\documentstyle[11pt,newpasp,twoside]{article}
\begin{document}
\title{$H_0$ from Gravitational Lenses: Recent Results}
\author{Paul L. Schechter}
\affil{Massachusetts Institute of Technology, 77 Massachusetts Avenue,
Cambridge, MA 02139-4307, USA}
\begin{abstract}
The 0$^{th}$, 1$^{st}$ and 2$^{nd}$ derivatives of a ``Fermat
potential" give the three D's of gravitational lensing: delay, deflection
and distortion.  Observations of these delays, deflections and
distortions for doubly and quadruply imaged quasars give estimates of
Hubble's constant, $H_0$.  The single largest contribution to the
uncertainty in $H_0$ arises from the difficulty in constraining the
degree of central concentration of the lensing potential.
Fortunately, astronomers have spent a good deal of effort over the
past quarter century addressing just this question.  If galaxies at $z=0.5$
are like nearby galaxies, the associated systematic uncertainty in
$H_0$ is less than 10\%.  The expected lens-to-lens scatter is 20\%.
Results from three particularly well constrained systems are
reported.
\end{abstract}
\section{Introduction}
Refsdal's (1964) method for measuring the Hubble constant using time
delays of gravitationally lensed quasars avoids all of the systematic
errors described by {\sc Freedman, Feast, Gibson, {\rm and} Dressler}
in their discussion of Cepheid based measurements: small parallaxes,
metallicity variations, reddening, crowding, blending, detector
non-linearities, unmodelled effects in secondary distance indicators,
and local deviations from pure Hubble flow.  Instead Refsdal's method
has systematic errors of its own, of which the best that can be said
is that they are completely different.

In what follows we first give a brief of outline\footnote{Narayan and
Bartelmann (1999) give an excellent and thorough introduction to
gravitational lensing.  The paper by Blandford and Kundi\'c (1997) on
the measurement of $H_0$ is likewise highly recommended.} of
gravitational lensing.  We report recent measurements of time delays
and discuss the uncertainties associated with predicting time delays
from models of gravitational potentials.  Using the three systems for
which these predictions are least uncertain, we report a result for
$H_0$.
\section{Gravitational Lensing}
In the weak field limit, gravitational potentials introduce an
effective index of refraction $n$ (e.g. Binney and Merrifield 1998),
which increases the light travel time for photons traveling through a
gravitational potential, $\Phi_{3D}$, 
\begin{equation} 
n = 1 - {2 \Phi_{3D} \over c^2 } \quad .
\end{equation}
A photon considering travel from a quasar at the edge of the universe
to the Milky Way will therefore plan on detouring around mass
concentrations if it wishes, consistent with Fermat's principle, to
take the path which takes the least time.  In the thin lens
approximation, the extra travel time $\tau$ will be proportional to a
dimensionless projected two-dimensional potential,
\begin{equation}
\psi_{2D} = {D_{LS} \over D_S} \int_{observer}^{source} {2 \Phi_{3D}\over c^2 } {d \ell \over D_L}
\quad ,
\end{equation}
where $D_{LS}$ and $D_S$ are angular diameter distances to the source
from the lens and the observer, respectively.  The extra travel
time is given by
\begin{equation}
\tau = {1+z_L \over c} {D_L D_S \over D_{LS}} \left[{1\over2}(\vec \theta - \vec \beta)^2
- \psi_{2D}(\vec \theta)\right] \quad ,
\end{equation}
where $\vec \beta$ is the angular position of the source and
$\vec \theta$ is the angular position at which the path intersects the
plane of the lens. Fermat's principle tells us that images form at
stationary points (minima, maxima and saddle points) of the extra travel
time.  These can be found by setting
$ {\partial \tau / \partial \vec \theta} = 0 $,
giving the famous ``lens equation'',
\begin{equation}
\quad\vec \theta - \vec \beta - {\partial \psi_{2D} \over \partial \vec \theta} = 0 \quad ,
\end{equation}
the solutions of which we denote, $\vec \theta_A$, $\vec \theta_B$, {\it et cetera}.
A small displacement of the source produces a small displacement of the image.
The ratio of these is the inverse magnification.  As both quantities are
vectors, differentiating the lens equation with respect to $\vec \theta$
gives the inverse magnification matrix:
\begin{equation}
{\partial \vec \beta \over \partial \vec \theta} =
\Biggl(
{1 - {\partial^2 \psi_{2D} \over \partial \theta_x^2}  
\quad - {\partial^2 \psi_{2D} \over \partial \theta_x \partial \theta_y}
\atop
- {\partial^2 \psi_{2D} \over \partial \theta_x \partial \theta_y} 
\quad 1 - {\partial^2 \psi_{2D} \over \partial \theta_y^2}}
\Biggr) \quad ,
\end{equation}
which is symmetric and therefore can be diagonalized.  Its inverse,
the magnification matrix, maps a circular source of unit area into an
elliptical image whose area is given by the absolute value of its
determinant.  Saddle points have one negative eigenvalue and maxima
have two.  Saddle points produce images with reversed handedness.

Taking the projected potential $\psi_{2D}$ and the source positions
$\vec \beta$ to be known, we can solve equation (4) for the image
positions $\vec \theta_A$ and $\vec \theta_B$ and obtain $\tau_A -
\tau_B$ from equation (3).  This difference depends on the difference
of the square bracketed dimensionless delays and some angular
diameter distances, which depend upon the Hubble constant and the
redshifts of the lens and source.  

All that is needed to determine the Hubble constant is a) a
measurement of the difference in travel time and b) a
model for the two dimensional potential which can be determined, in
principal, from the deflections and distortions of the observed
images, and from any surplus delays.  We shall see that in practice
there are daunting challenges both in the measurement of delays and in
the modelling of potentials.

\section{Time Delay Measurements}

The quantities outside the square brackets in equation (3) can be
rewritten as a dimensionless quantity (of order unity for quasars)
divided by $H_0$.  The scale of the square bracketed quantity is set
by the square of the image deflection, $\vec \theta - \vec \beta$, which for
lensed quasars is of order $3 \times 10^{-6}$ radians.  The typical
difference in travel time is therefore of order $10^{-11}/H_0$
and is measured in weeks or months.

This is at best a mediocre match to the intrinsic variability
timescale for quasars, which is of order several years (e.g. Cristiani
et al. 1996).  To make matters worse, the typical variation on that typical
timescale is only of order 10\%, with smaller variations on shorter
timescales.  One therefore needs high photometric or radiometric
accuracy to measure a time delay.  It is somewhat misleading to talk
of ``typical'' variability, since some quasars (e.g. 3C279 and BL Lac)
are atypically variable while many others, to the chagrin of those who
would measure time delays (e.g. Saust 1991, Moore and Hewitt 1997), are atypically
steady.

Lens delay timescales are also poorly matched to more local phenomena.
Delays of 2 weeks are difficult to measure in the optical if the
object can only be observed during the dark run.  Delays of 6 months
are difficult to measure in the optical if the object lies near the
equator.  Delays of more than 6 months can be difficult to measure
with the VLA because it changes configuration.

Yet another obstacle is microlensing by the stars in the intervening
galaxy (Chang and Refsdal 1979).  The rms amplitude for microlensing
can be as high as 0.5-1.0 magnitudes for an unresolved source (Witt et
al. 1995).  Predictions of microlensing on a timescale of years for
source velocities of a few hundred km s$^{-1}$ have been dramatically
confirmed (Wozniak et al. 2000).  But microlensing on a timescale of
days has now been observed both at optical (Burud et al. 2000) and
radio wavelengths (Koopmans and de Bruyn 2000), and may be more
general than has heretofore been appreciated.  While microlensing may
be signal for investigators who want to measure the mass fraction in
compact objects, it is noise for those who wish to measure time
delays.

\begin{table}
\caption{\bf Measured time delays}
\begin{tabular}{lcl}
\tableline
Lens & Delay & Investigator \\
\tableline
B0957+561& $417^{\rm d} \pm \ 3$ & Kundi\'c {et al.} 1997 \\
         & $403^{\rm d} \pm 30$  & Haarsma {et al.} 1999 \\
PG1115+080* & $\ 25\fd 0 \pm 1.6$  & Schechter {et al.} 1997 \\
            & & Barkana 1997  \\
B1608+656*  & $73^{\rm d} \pm 3$ & Fassnacht { et al.} 1999, 2000 \\
B0218+357   & $\ 10\fd5 \pm 0.3$ & Biggs { et al.} 1999 \\
PKS1830-211& $\ \ \, 26^{\rm d} \pm 4.5$ & Lovell { et al.} 1998 \\
           & $\ \ 24^{\rm d} \pm 6 \ \  $ & Wiklind \& Combes 1999 \\
HE1104-1805& $ 267^{\rm d} \pm 90$     & Wisotzki { et al.} 1998  \\
B1600+434  &$ 47^{\rm d} \pm 5$ & Koopmans { et al.} 2000  \\
           &$ 51^{\rm d} \pm 2$ & Burud { et al.} 2000a \\
RXJ0911+0551 & $ 200^{\rm d} \pm 40$ & Burud { et al.} 2000b\dag \\
\tableline
\tableline
\end{tabular}
\break
* Multiple delays measured; longest reported. \hfill
\break
\dag \ Author's handwritten notes, 1999 Boston University lens conference.
\end{table}

There are also sociological hurdles.  An optical monitoring campaign
might require one 15 minute observation every other night for 3 months,
for a total 11 hours of observing time.  Instrument changes and
scheduling policies make such programs exceedingly difficult, even at
observatories which attempt to accommodate them.  Try entering a
fraction of an hour in the ESO proposal form on the ``time requested''
line -- the number will be rejected!  Successful monitoring campaigns
have relied on a combination of pleading, arm-twisting and
horsetrading.  They also produce substantial telephone bills.

These difficulties notwithstanding, time delays have now been measured
for 8 systems, as summarized in Table 1.  The efforts reported there
have been truly heroic, with the observers having gone to
extraordinary lengths to produce accurate photometry and radiometry.
Of particular note is the effort reported by {\sc Fassnacht et al.,}
at this meeting, who have measured all three independent time delays
for the system CLASS 1608+656.  Considerable good work has also gone
into the difficult measurement of redshifts for the lensing galaxies.
But it is in the nature of the enterprise that we spend most of our
time on the bad news rather than the good, so we turn to modelling
potentials.

\section{Modelling Potentials}
One can, with little effort, find very different values for $H_0$ in
the literature based on the same measured time delay.  Not
surprisingly, the differences can be traced to differences in the
model for the lens potential.  Unfortunately, models are unavoidable,
since lens potentials can never be determined with arbitrary accuracy.
For the sake of consistency we adopt here a simple ``yardstick'' model
which one may then compare with more refined models as the data
permit.  The yardstick also suffices to illustrate the principal
systematic errors associated with model predictions.

\subsection{A ``yardstick'' model}
We take the lensing galaxy to be circularly symmetric in projection on
the sky, with a two dimensional potential given by
\begin{equation}
\psi_{gal} = {b^2 \over (1 + \alpha)} \left( {\theta \over b} \right)^{1+\alpha}
\end{equation}
where $b$ gives the angular size of the ``Einstein ring'' that the lens would
produce for a source on axis.  The singular isothermal sphere corresponds
to $\alpha = 0$, in which case $b = {4 \pi \sigma^2 / c^2}$, where $\sigma$
is the one dimensional velocity dispersion.  For sources close to the axis
(Refsdal and Surdej 1994) we have
\begin{equation}
\tau_B - \tau_A \approx {1+z_L \over H_0} {d_L d_S \over d_{LS}} 
{1\over 2}(\theta_A^2 - \theta_B^2)(1 - \alpha) \quad ,
\end{equation}
where the dimensionless parts of the angular diameter distances are
given by $d = D H_0 / c$.  Witt, Mao and Keeton (2000) have shown
that strict equality holds for $\alpha = 0$ and for related
self-similar potentials.

An important lesson to be learned from equation (7) is that one cannot
tolerate much uncertainty in the center of the lensing potential,
particular for images which are nearly equidistant.  This has unhappy
consequences for several of the lenses for which time delays are
reported in Table 1.

In two cases the lensing galaxy is barely visible or not at all,
making it difficult to determine where its center lies.  In another
the lens appears compound, comprised of merging components.  It seems
risky, for such a lens, to assume that the mass (in particular the
dark matter) is comprised of components which are centered on the
light.  While such close pairs might at first seem unlikely, there are
other cases known (Wisotzki et al. 1999, Rusin et al. 2000).  A close
pair of galaxies will have a larger quadrupole moment than a single
galaxy, and is therefore more likely to give a highly magnified and
therefore overrepresented (Turner et al.  1984) quadruple image
configuration.

But we've gotten ahead of ourselves.  The potential of equation (6)
will only produce pairs of images, or occasionally triples if $\alpha
> 0$ (Rusin and Ma 2000).  Quadruple systems, which
are frequently seen, require that the potential have a quadrupole,
possibly attributable to a tide,
\begin{equation}
\psi_{tide} = {\gamma \over 2} \theta^2  \cos 2(\phi - \phi_\gamma) \quad,
\end{equation}
where $\gamma$ (sometimes called the shear) measures the strength of
the tide and $\phi_\gamma$ measures its orientation on the plane of
the sky.  If the observed quadrupole is strong, one can, assuming a
tidal origin, look for higher order terms and solve for the position
of the tidal perturber.  In the cases of PG1115+080, RXJ0911+0551,
B0957+561, B1422+231, CLASS 2045+265 and B1413+113 a group or cluster
of galaxies is observed with its center at the computed position and
with a measured or plausible redshift identical to that of the lensing
galaxy.

The sense of satisfaction one gets from identifying the group of
galaxies responsible for the tide quickly becomes dismay when one
realizes that the group extends beyond the lens and therefore projects
a mass surface density onto it.  A mass sheet of uniform density
$\Sigma$ produces a projected potential
\begin{equation}
\psi_{sheet} = \kappa \theta^2 \quad ,
\end{equation}
where the dimensionless surface density (sometimes called the
convergence) is given by $\kappa = 4 \pi G \Sigma D_L D_{LS} / D_{S}
c^2 $.  Adding a mass sheet to a lens produces an image configuration
which is consistent with a scaled version of the original lens
potential.  The time delays predicted by this scaled potential are
longer than those predicted by the actual (lens plus sheet) potential
by a factor of $1/(1-\kappa)$.  But unless one knows the intrinsic
size (or luminosity) of the lensed object, one cannot tell from the
observed deflections, distortions, and delays whether a mass sheet is
present or not (Saha 2000).  This ``mass sheet degeneracy'' makes it
impossibile to predict a time delay unambiguously.

Different investigators have chosen different approaches to this
problem.  One can estimate the mass surface density of the cluster
through measurements of its velocity dispersion (Angonin-Willaime et
al. 1994) or through weak lensing (Fischer et al. 1997).
Alternatively one can estimate the expected deflection due to the lens
alone through measurement of the velocity dispersion of the lensing
galaxy (Tonry and Franx 1999).  Absent such measurements, one can
still estimate the effect of such a cluster if one is willing to
assume that the cluster is isothermal.  One then finds that the
convergence is equal to the shear,
\begin{equation}
\kappa = \gamma  \quad {\rm (for\ an\ isothermal\ cluster).}
\end{equation}
Operationally, one finds the best model using equations (6) and (8)
and corrects for the effect of the projected mass sheet by multiplying
the predicted time delay by a factor $1-\gamma$.

\subsection{Constraining models$^2$}
\footnotetext[2]{The material in this subsection was
omitted in the 16 minutes allotted to the spoken version of this
review.}
Our yardstick model for the potential has four free parameters
associated with it: the lens strength, $b$, the tidal shear $\gamma$,
its orientation, $\phi_\gamma$ and the radial exponent of the
potential $\alpha$.  But there are free parameters associated with the
source as well.  For example the source position, $\vec \beta$ has two
components which cannot be directly observed and therefore must count
as model parameters.

\begin{table}
\caption{\bf Count of Source Parameters and Constraints}
\begin{tabular}{lcccl}
\tableline
observable & source     & constraints & constraints & fractional \\
           & parameters & (2 images)  & (4 images)  & accuracy   \\
\tableline
delay      & 1 & 1 & 3 & .01-.1 \\
position   & 2 & 4 & 8 & .001 \\
flux*	   & 1 & 2 & 4 & .01-.5 \\
shape*     & 3 & 6 & 12 & .1 \\
\tableline 
\tableline
\end{tabular}
\break
*Shape uses the full magnification matrix; flux uses only its determinant.
\end{table}

In Table 2 we show how different observable quantities contribute
additional source parameters and constraints.  We show delay
observations adding one free parameter, $H_0$.  We count ``shapes'' as
having a size, an axis ratio and an orientation.  Not all constraints
are equally useful.  Position constraints are typically accurate to
one part in a thousand.  Fluxes might be good to a part in one
hundred, but microlensing may make these uncertain by as much as 50\%.
Surplus delays (beyond the one needed to measure $H_0$) constrain the
projected potential.  Positions constrain its first derivatives.
Fluxes or shapes (one cannot use both simultaneously) constrain
second derivatives.

The simplest lenses, those with two unresolved images, present a
problem.  Measuring their positions and (possibly microlensed) fluxes
we have 4 parameters associated with the yardstick model and 3 source
parameters, and only 6 constraints.  At this point we introduce a
``prior''.  We know a good deal about the shapes of galaxy potentials from
studies of nearby galaxies.  These appear to be approximately
isothermal.  Short of discarding doubles, we may reasonably assume
that the lensing galaxies are isothermal and take $\alpha = 0$.

Extended sources can, in principal, be decomposed into a set of finite
sources, for each of which one can measure delays, deflections and
distortions.  In practice the success in using extended sources to
constrain models has been spotty (Chen et al. 1995; Kochanek et al.
2000) with the successes involving high surface brightness
sources.

Even without the benefit of Table 2, it is clear that systems with
more images and more sources are better than those with fewer.  But a
system with fewer constraints for which one suspects that the
potential is relatively simple may be more useful than one for which
the potential is likely to be complicated.  CLASS 1359+154 has one source
with six images, but the lens is a triplet of galaxies.  How would one
expect the dark matter to be distributed such a system?

\subsection{Radial exponent: the return of the mass sheet degeneracy}

Perhaps the largest source of variance among estimates of $H_0$ which
use the same underlying observations can be traced to differing
treatments of the radial exponent $\alpha$.  Almost universally,
investigators have found it exceedingly difficult to constrain
$\alpha$ and its equivalents.  The reader is referred to figure 4a in
Williams and Saha (2000), figure 3 in Bernstein and Fischer (1999) and
figure 7 in Cohn et al. (2000), which show the wide the range of
allowable values of $\alpha$ and the strong dependence of $H_0$
thereon.  For the case of the quadruple PG1115+080, the difference
between the best fitting isothermal model ($\alpha = 0$) and the best
fitting point mass model ($\alpha = -1$) amounts to only several parts
in a thousand in the positions of the images.

Over the relatively narrow range of radii over which one is likely
to see four images, a change in power law is not very different from
adding (or subtracting) a mass sheet.  Even in the case of 1933+503,
with two quadruply imaged sources and one doubly imaged source,
Cohn et al. (2000) find it difficult to constrain the concentration.

At this point the pragmatist will return to a prior based on the
accumulated knowledge of nearby galaxies.  Studies of spiral galaxies
show their rotation curves to be very nearly flat, corresponding to
$\alpha = 0$.  Studies of the line of sight velocity distribution as a
function of radius in ellipticals have yielded slightly steeper
potentials.  Romanowsky and Kochanek (1999) have collected and
averaged these results.  Application of a ruler to their figure 1
yields $<\alpha> = -0.2$, with a galaxy-to-galaxy scatter of roughly
0.2.

For a handful of lenses (only one of which has a measured time delay)
the image configuration does suffice to constrain $\alpha$ (Chen and
Kochanek 1995; Barkana et al. 1999; Cohn et al. 2000).  In all of
cases the results are consistent with the isothermal hypothesis, with
uncertainties in $\alpha$ of 0.1-0.2.  Rusin and Ma (2000) note that
the absence of third images restricts $\alpha$ to values $\la 0.1$. 

As most observed lensing galaxies appear to be elliptical (Keeton,
Kochanek and Falco 1998) we shall proceed under the assumption that
$<\alpha> = -0.2$, with a guess at the uncertainty in the mean of 0.1,
and a corresponding systematic uncertainty in $H_0$ of 10\%.  Given
the observed scatter in $\alpha$ of 0.2 in nearby ellipticals, we
expect scatter in individual values of $H_0$ of 20\%.

\section{The Hubble Constant}

At the risk of deeply offending some large fraction of the local
organizing committee we shall not apply our ``yardstick'' model to all
the lens systems for which delays have been measured.  In particular
we shall set aside a) those systems for which the center of the
lensing galaxy cannot be reliably identified from optical/IR images
and b) those systems for which there are only two unresolved images,
and which therefore depend on magnifications to constrain the model.
For both unresolved doubles, B1600+434 and HE1104-1805, there is
substantial evidence that microlensing affects the fluxes.  The case
of B1600+434 is further complicated by the possible effects of the
bright galaxy which lies just a few arcseconds from the system.

There is considerable danger in deciding {\it post hoc} where to draw
the line on including or not including systems.  While other
investigators have chosen not to exclude those systems excluded here,
I suspect there is general agreement on the relative reliability of
the predicted time delays for the lenses in Table 1.

\begin{table}
\caption{\bf Predicted and observed delays}
\begin{tabular}{lllll}
\tableline
system & $\gamma$ & $\Delta \tau_{pred}$ & $\Delta \tau_{obs}$ & $h$ \\
\tableline
 B0957+561  & 0.279 & $274^{\rm d}/h$ & $417^{\rm d}$ & 0.66 \\
 PG1115+080 & 0.110 & $10\fd 2/h$     & $25\fd 0$     & 0.41 \\
 RXJ0911+0551 & 0.307 & $83\fd 5/h$   & $200^{\rm d}$ & 0.42 \\
\tableline 
 average & & &  $<h> = $ & 0.494 \\
\tableline
 $\times <1 - \alpha>$ & & & $<h> = $ & 0.592 \\
\tableline
 $ (\Omega_m,\Omega_\Lambda) = (0.3,0.7)$ & & & $<h> = $ & 0.622 \\
\tableline 
\tableline
\end{tabular}
\end{table}

Applying our isothermal yardstick model to the three remaining systems
gives the predicted time delays in Table 3.  The individual
predictions have had the factor $1 - \gamma$ applied on the assumption
that the cluster responsible for the shear is also isothermal with
$\kappa = \gamma$.  Defining $h \equiv H_0 / (100 {\rm km\ s}^{-1}
{\rm Mpc}^{-1})$, its average value is then corrected for the assumed
deviation of the lensing galaxy from isothermality.  Our angular
diameter distances, $D$, were computed using an Einstein-deSitter
model.  Adopting $\Omega_m = 0.3$ and $\Omega_\Lambda = 0.7$ gives
another small correction.

From the variations in $\alpha$ in nearby galaxies we expect a
dispersion in the derived values of $h$ of 20\%.  The uncertainties in
the measured time delays add (in quadrature) as much as 20\%.  If you
want to bet on the statistics of three objects, it is straightforward
to derive a statistical uncertainty in $h$.

There are two paths to a better Hubble constant, both of which require
finding new lensed systems.  The purists seek a few ``golden'' lenses
which are well constrained by the image configuration.  The
pragmatists are willing to work with less well constrained systems,
and are willing to take advantage of ``prior'' information gleaned
from nearby galaxies.

Astronomers in both camps can take heart in the fact that there are
many more bright lenses waiting to be discovered.  By way of
illustration, there are some 5500 quasars brighter than $B=18.5$ in
the 7500 square degrees of the Hamburg-ESO survey (Wisotzki et al.
2000).  Imaging these at high resolution should yield (in addition to
the six already found) roughly 18 as yet undiscovered lenses for a
conservative, Einstein-deSitter model.  Whether by brute force or
elegance, gravitational lenses will give a competitive value for $H_0$.

\acknowledgements
This work was supported in part by the U.S. National Science
Foundation, grant AST96-16866.

\end{document}